# DG-Embedded Radial Distribution System Planning Using Binary-Selective PSO

Ahvand Jalali

University of Melbourne,

Melbourne, Australia

ajalali@student.unimelb.edu.au

S K. Mohammadi

Islamic Azad University,

Boukan Branch, Iran

s_kadkhoda63@yahoo.com

H. Sangrody

Binghamton University,

New York, USA

habdoll1@binghamton.edu

A. Rahim-Zadegan

Karlsruhe Institute of Technology,

Karlsruhe, Germany

aso.rahimzadegan@student.kit.edu

*Abstract* –**With the increasing rate of power consumption, many new distribution systems need to be constructed to accommodate connecting the new consumers to the power grid. On the other hand, the increasing penetration of renewable distributed generation (DG) resources into the distribution systems and the necessity of optimally place them in the network can dramatically change the problem of distribution system planning and design. In this paper, the problem of optimal distribution system planning including conductor sizing, DG placement, alongside with placement and sizing of shunt capacitors is studied. A new Binary-Selective Particle Swarm Optimization (PSO) approach which is capable of handling all types of continuous, binary and selective variables, simultaneously, is proposed to solve the optimization problem of distribution system planning. The objective of the problem is to minimize the system costs. Load growth rate, cost of energy, cost of power, and inflation rate are all taken into account. The efficacy of the proposed method is tested on a 26-bus distribution system.**

*Index Terms*- **Distribution system planning, conductor sizing, capacitor placement, DG placement, binary-selective PSO.**

## I. Introduction

Distribution system, responsible for transferring electrical energy to the end users, plays a determining role in the power system economics. Operating at low voltages, and high currents, it suffers from high power loss which is considered as a continuous cost for the distribution system and relates reversely with the size of the system conductors. Besides, keeping the voltage profile and current flows of the system in the acceptable operating range constrains the problem of conductor sizing since the small conductors have less current flow capability and also lead to more voltage drops across the system. This necessitates an economic viewpoint in selecting the system conductors. The problem of conductor sizing is addressed in many papers using a variety of methods including analytical [1], evolutionary [2, 3], and other creative approaches [4, 5]. The common objective for this problem has been to minimize the system costs which include loss cost and conductors cost. Maintaining system's voltages and currents within their desired margins are also the prevalent restrictions for such problem.

In addition, installing shunt capacitors is widely used for power flow control, power factor correction, voltage profile management and losses reduction [6]. Again, an optimal placement and sizing of capacitors is necessary to make the most of their capability. Shunt capacitor placement problem has also been addressed in the literature using heuristic approaches [7, 8], sensitivity analysis [8], fuzzy [9], etc.

With the increasing penetration of renewable DGs into the power systems, optimal placement of DG units in the distribution network is becoming more important [10]. DG placement has been addressed in [11] for the purpose of minimizing the consumer's cost.

Supplying active and reactive loads locally, DGs and capacitors alter the power flow of the system; hence, play the same role as the supplementary conductors. Thus, both problems of DG and capacitor sizing and placement are of close correlation with the conductor sizing problem. If taking into account all these problems simultaneously, more desirable results, i.e. less investment cost and more loss reduction, can be achieved.

Particle swarm optimization is proved to be one of the most computation-efficient heuristic approaches. Ref. [12] has shown that PSO, compared to GA and conventional methods, has a better performance on capacitor placement problem. Binary versions of the PSO has also been developed [13, 18] which are of high advantages for solving optimization problems such as DG and capacitor placement.

The combination of conductor sizing and capacitor placement has been addressed in several references using PSO [14], GA [15] and other creative methods [16]. In this paper, a sub-problem of renewable DG placement is also included in the problem. Furthermore, the multi-objective optimization problem of conductor sizing is solved with the weighting factors being assigned to the elements of the objective function and meaningful conductor profiles are obtained. A new PSO procedure is proposed which is capable of choosing the variables from a selective space. This is useful since the available conductors, capacitors and DGs are normally of standard sizes which the optimizer should select the optimal ones. A typical 26-bus distribution system is utilized to show the practicality of the proposed method.

## II. Problem Formulation

The objective function of the problem is defined in this paper as the summation of two main cost components as defined in (1). The first cost component is the capital cost of the conductors which is incurred to the system at the beginning of the planning period. The other is the cost of the lost power and energy which is imposed on the system continuously during the entire planning period.

$$OBJ = \sum_{k=1}^{N_k} Cond\_Cost_k + \sum_{t=1}^{T} Loss\_Cost_t \quad (1)$$

In (1), $k$ is the set of system sections, $N_k$ is the total number of sections, $t$ is the set of planning horizon years, and $T$ is the total years of the planning horizon. Also, $Cond\_Cost_k$ and $Loss\_Cost_t$ are the capital cost of the conductor of section $k$ and the cost of lost power and lost energy in year $t$ ($), as defined in (2) and (3), respectively:

$$Cond\_Cost_k = C\_Cost_k \times L_k \quad (2)$$

$$Loss\_Cost_t = Ploss_t \times [CP_t + (CE_t \times LsF \times 8760)] \quad (3)$$

where $C\_Cost_k$ is the price per kilometer of conductor for section $k$ ($/km), $L_k$ is the length of section $k$ (km), $Ploss_t$ is the active power loss in year $t$ (kW); and $CP_t$ and $CE_t$ are cost of power and energy in year $t$ ($/kW, $/kWh), respectively. Also, $LsF$ is the loss factor which, according to the British experience, can be expressed in terms of the load factor, $LF$, i.e. the ratio of the average load to the peak load, as (4) [1]:

$$LsF = 0.2 \times LF + 0.8 \times LF^2 \quad (4)$$

For each year of the planning period, considering the effect of the inflation rate ($inf$), $CP_t$ and $CE_t$ are calculated as (5) and (6), respectively:

$$CP_t = CP_0 \times (1 + inf)^t \quad (5)$$

$$CE_t = CE_0 \times (1 + inf)^t \quad (6)$$

Moreover, $Ploss_t$ in (3) can be found through the difference between total power generation and consumption in year $t$ as (7):

$$Ploss_t = P_s^t + \sum_{j=1}^{N_j}(M_j \times P_{DG\,j}^t) - \sum_{j=1}^{N_j} P_{Dj}^t \quad (7)$$

where $P_s^t$ is the power supplied from the upstream network, $M_j$ is the binary variable implying DG placed at bus $j$, and $P_{DG\,j}^t$ is the active power generated by DG at bus $j$ and year $t$ (kW). The active power demand in bus $j$ and year $t$, $P^t_{Dj}$, is defined as (8) considering the effect of the load growth rate ($LR$). The case is similar for reactive demand, $Q^t_{Dj}$.

$$P_{Dj}^t = P_{Dj}^0 \times (1 + LR)^t \quad (8)$$

The main constraints of the problem are active and reactive power balance in each node and in each planning horizon year, as (9) and (10). All variables with superscript $t$ are associated with year $t$ of the planning horizon.

$$P_s^t + (M_j \times P_{DG\,j}^t) - P_{Dj}^t =$$
$$U_j^t \sum_{m=1}^{Nj} [U_m^t \{G_k cos(\delta_j^t - \delta_m^t) + B_k sin(\delta_j^t - \delta_m^t)\}] \;\forall j, t \quad (9)$$

$$Q_s^t + (M_j \times Q_{DG,j}^t) + Qc_j - Q_{Dj}^t =$$
$$U_j^t \sum_{m=1}^{Nj} [U_m^t \{G_k sin(\delta_j^t - \delta_m^t) - B_k cos(\delta_j^t - \delta_m^t)\}] \;\forall j, t \quad (10)$$

where $U_j^t$ and $\delta_j^t$ are the voltage magnitude and phase angle of bus $j$ in year $t$, respectively; and $Qc_j$ is the capacitor's reactive power at bus $j$ specified via the capacitor type selected for bus $j$ by the selective PSO algorithm. Also, the conductance, $G_k$, and susceptance, $B_k$, of section $k$ are determined through the resistance, $R_k$, and reactance, $X_k$, of the chosen conductor type for section $k$ by the selective PSO optimization method, as will be explained in section III, and are calculated using (11) and (12), respectively:

$$G_k = Real[1/(L_k \times (R_k + jX_k))] \quad (11)$$

$$B_k = Imag.[1/(L_k \times (R_k + jX_k))] \quad (12)$$

Other inequality constraints are the allowable limits for all the $k^{th}$ feeder's current, $I_k^t$, and bus voltages, as (13) and (14), respectively:

$$-I_k^{max} < I_k^t < I_k^{max} \quad (13)$$

$$U^{min} < U_j^t < U^{max} \quad (14)$$

where $I_k^{max}$, the maximum current capability of the conductor of section $k$, is also determined through the conductor type selected for the section $k$ by the selective PSO. Finally, $U^{min}$ and $U^{max}$ are set at 0.95 p.u and 1 p.u, respectively.

Another cost component of the problem is capacitors' cost including capital and installation cost as (15):

$$Cap\_Cost = \sum_{j=1}^{Nj}(Cptl_{C\,j} + Inst_{C\,j}) \quad (15)$$

Both $Cptl\_C_j$ and $Inst\_C_j$, i.e. the capital and installation costs of capacitor at bus $j$, are determined via the capacitor type selected for bus $j$ by the selective PSO algorithm. The capacitor placement cost is included in the problem through defining the following constraint on the maximum budget for the capacitor, i.e. $Cap\_Cost_{max}$, as (16):

$$Cap\_Cost < Cap\_Cost_{max} \quad (16)$$

Similarly, DG placement is incorporated into the problem through equations (17) and (18):

$$DG\_Cost < DG\_Cost_{max} \quad (17)$$

$$DG\_Cost = \sum_{j=1}^{N_j} M_j \times (Cptl\_DG + Inst\_DG) \qquad (18)$$

where $DG\_Cost$ and $DG\_Cost_{max}$ are total cost and maximum budget for DGs, respectively. Besides, $Cptl\_DG$ and $Inst\_DG$ are capital and installation cost of the available DG type. As seen, unlike capacitor, the DG selection at bus $j$ is decided via the binary variable $M_j$. It is to fully demonstrate the capability of the proposed PSO algorithm in handling both types of binary and selective variables. In this paper, one DG type is supposed to be available for installation without loss of generality.

### III. BINARY-SELECTIVE PSO PROCEDURE

Particle Swarm Optimization (PSO) is a search-based optimization method, based on movement and intelligence of swarms; in which individuals orient their movements towards the personal (*pbest_i*) and overall (*gbest*) best locations determined by calculation of the associated objective function of particles [17]. The modification of particles is defined through the velocity concept as (19):

$$v_i^{it+1} = w^{it} v_i^{it} + c_1 \times rand \times (pbest_i - s_i^{it}) + c_2 \times rand \times (gbest - s_i^{it}) \qquad (19)$$

where $it$ is the set of iterations of PSO, $v_i^{it}$ and $s_i^{it}$ are the velocity and position of agent $i$ at iteration $it$, respectively, $c_{1,2}$ are constants, $rand$ is a random number between 0 and 1, and $w^{it}$ is the weighting factor at iteration $it$ calculated as (20):

$$w^{it} = w_{max} - [(w_{max} - w_{min})/it_{max}] \times it \qquad (20)$$

According to the examinations carried out in [17], the values $c_1 = c_2 = 2.0$, $w_{max} = 0.9$, and $w_{min} = 0.4$ have been realized to be proper, independent of the problem. In the Binary version of PSO (BPSO), the next state of each particle is determined according to the agent's tendency to be zero or one. The higher the velocity, the more likely the agent to choose 1, vice versa; as (21) and (22) [18]:

$$sigmoid(v_i^{it+1}) = \frac{1}{1 + exp(-v_i^{it+1})} \qquad (21)$$

$$s_i^{it+1} = \begin{cases} 1 & if \ rand < sigmoid(v_i^{it+1}) \\ 0 & otherwise \end{cases} \qquad (22)$$

As mentioned in section II, in this paper DG placement is included in the problem through the binary variable $M_j$ which is decided through (21) and (22).

In many engineering problems, the capability of the optimizer to select the decision variables from a set of available, standard values is required. In this paper, the binary PSO formulation is modified to give it the capability of searching in a selected space (selective PSO) as follows. Derived from the same concept of (22), the higher value of the agent's velocity indicates its inclination to take higher values in the selective space which in the current problem means to pick up higher conductor and capacitor sizes. This concept can be modeled for the conductor sizes as (23):

$$s_i^{it+1} = \qquad (23)$$

$$\begin{cases} S_1(C\_Cost^1, R^1, X^1, I^{max1}) & rd + a_1 < sigmoid(v_i^{it+1}) \\ S_2(C\_Cost^2, R^2, X^2, I^{max2}) & rd + a_2 < sigmoid(v_i^{it+1}) \\ \vdots \\ S_n(C\_Cost^n, R^n, X^n, I^{maxn}) & rd + a_n < sigmoid(v_i^{it+1}) \end{cases}$$

where $C\_Cost^1$, $R^1$, $X^1$, and $I^{max1}$ are conductor cost, resistance, reactance (all per-kilometer), and current carrying capability of the conductor *type 1*, respectively; and similarly for the other conductor types. Besides:

$$-0.5 \leq a_n < \ldots < a_2 < a_1 \leq 0.5 \qquad (24)$$

This makes the algorithm select larger conductor sizes for particles with higher velocity and thus higher sigmoid function, vice versa. It is supposed that $S_1$ is the largest available conductor size and the *types 2* to *type n* are getting smaller in size step by step.

Similarly for the capacitor sizes, the model is as (25):

$$s_i^{it+1} = \qquad (25)$$

$$\begin{cases} C_1(Qc^1, Cptl\_C^1, Inst\_C^1) & rd + \beta_1 < sigmoid(v_i^{it+1}) \\ C_2(Qc^2, Cptl\_C^2, Inst\_C^2) & rd + \beta_2 < sigmoid(v_i^{it+1}) \\ \vdots \\ C_n(Qc^n, Cptl\_C^n, Inst\_C^n) & rd + \beta_n < sigmoid(v_i^{it+1}) \end{cases}$$

where $Qc^1$, $Cptl\_C^1$, and $Inst\_C^1$ are reactive power capability, capital cost, and installation cost of the capacitor *type 1*, respectively; and similarly for the other capacitor types. Again, $C_1$ indicates the largest available capacitor type and the other types are getting smaller successively. Similarly:

$$-0.5 \leq \beta_n < \ldots < \beta_2 < \beta_1 \leq 0.5 \qquad (26)$$

The last capacitor size should be defined zero, implying no capacitor selected for the associated bus. Inasmuch as the problem contains both binary and selective variables to be optimized, all equations (19) to (26) are used at the same time to update the associated particles' position. The proposed version of PSO can be referred to as Binary-Selective PSO or BSPSO.

To ensure the feasibility of each modified particle, it must be checked to satisfy all the problem constraints. If any constraint is violated, the particle should be disposed of. This can be easily carried out via assigning an extreme value to the particle's associated objective function. Thus, it won't be selected as *pbest_i* or *gbest* and will be removed from the cycle of iterations.

## IV. SIMULATION RESULTS AND DISCUSSION

The proposed method has been tested on a 26-bus distribution test system. Table I shows the information of the system configuration and loads. The single-line diagram of the system is also shown in Fig. 1. The system rated values are 20kV and 1 MVA. The cost of power and energy are assumed 168 $/kW and 0.06 $/kWh, respectively [19]. Besides, an inflation rate and load growth rate as 0.05 and 0.02, respectively, are considered. The system has the load factor as 0.25 and the power factor of all loads is assumed to be 0.85 lagging. The problem is investigated for a 10-year planning period.

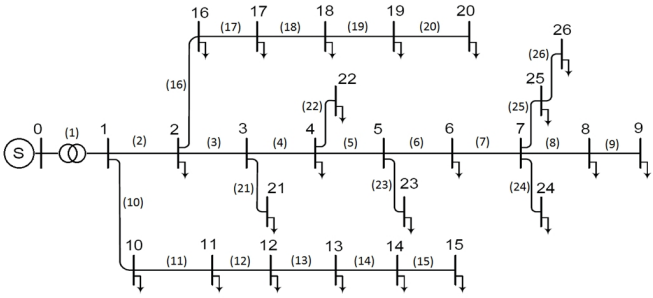

Figure. 1 The 26-bus distribution test system

TABLE I. DISTRIBUTION SYSTEM DATA

| Section (k) | From bus | To bus | Length (km) | $S_{load}$ (kVA) | Section (k) | From bus | To bus | Length (km) | $S_{load}$ (kVA) |
|---|---|---|---|---|---|---|---|---|---|
| 1 | 0 | 1 | 0.000 | 0 | 14 | 13 | 14 | 2.625 | 725 |
| 2 | 1 | 2 | 1.175 | 0 | 15 | 14 | 15 | 2.925 | 900 |
| 3 | 2 | 3 | 0.625 | 950 | 16 | 2 | 16 | 1.175 | 300 |
| 4 | 3 | 4 | 1.825 | 0 | 17 | 16 | 17 | 0.650 | 750 |
| 5 | 4 | 5 | 0.850 | 0 | 18 | 17 | 18 | 1.825 | 350 |
| 6 | 5 | 6 | 1.125 | 850 | 19 | 18 | 19 | 0.825 | 400 |
| 7 | 6 | 7 | 2.625 | 0 | 20 | 19 | 20 | 1.125 | 700 |
| 8 | 7 | 8 | 2.925 | 640 | 21 | 3 | 21 | 2.625 | 125 |
| 9 | 8 | 9 | 1.175 | 813 | 22 | 4 | 22 | 2.925 | 565 |
| 10 | 1 | 10 | 0.650 | 800 | 23 | 5 | 23 | 1.175 | 682 |
| 11 | 10 | 11 | 1.825 | 400 | 24 | 7 | 24 | 0.650 | 900 |
| 12 | 11 | 12 | 0.825 | 950 | 25 | 7 | 25 | 1.825 | 575 |
| 13 | 12 | 13 | 1.125 | 825 | 26 | 25 | 26 | 0.825 | 200 |

The information of the available conductor and capacitor types is given in tables II and III, respectively. The last defined type of the capacitors indicates no compensation in the associated bus. As many of the available renewable DG types such as inverter-interfaced DGs or Doubly Fed Induction Generator (DFIG) have reactive power capability, one DG unit type with the rated active and reactive generation capacity of 500 kW and 300 kVAr, respectively, and total capital and installation cost of $4000 is considered to be optimally placed in the system. Two scenarios have been implemented as follows:

TABLE II. AVAILABLE CONDUCTORS' DATA

| Conductor Type | R (Ω/km) | X (Ω/km) | Price ($/km) | $I_{max}$ (A) |
|---|---|---|---|---|
| 1 | 0.158 | 0.23 | 151 | 520 |
| 2 | 0.271 | 0.25 | 81 | 310 |
| 3 | 0.455 | 0.26 | 48 | 212 |
| 4 | 0.782 | 0.28 | 31 | 150 |
| 5 | 1.374 | 0.39 | 15 | 107 |

TABLE III. AVAILABLE CAPACITORS' DATA

| Capacitor Type | Size (kVAr) | Price ($) | Installation Cost ($) |
|---|---|---|---|
| 1 | 1200 | 2040 | 100 |
| 2 | 600 | 1320 | 100 |
| 3 | 300 | 975 | 100 |
| 4 | 0 | 0 | 0 |

***Scenario A.*** *Conductor sizing only:* In this scenario, only conductor sizing is focused with the objective function as (1) and neither capacitor nor DG is included in the problem. The results of this scenario are shown in Table IV.

The variable $U_{ind}$ is used as a criterion of the voltage profile smoothness and is defined as (27):

$$U_{ind} = \sum_{j=1}^{N_j} abs(1 - U_j) \qquad (27)$$

TABLE IV. THE RESULTS OF CONDUCTOR SIZING FOR CASE 1

| Section (k) | Chosen Cond. | U (p.u) (ending bus) | Section (k) | Chosen Cond. | U (p.u) (ending bus) | Section (k) | Chosen Cond. | U (p.u) (ending bus) |
|---|---|---|---|---|---|---|---|---|
| 1 | - | 1 | 10 | 1 | 0.99 | 19 | 3 | 0.98 |
| 2 | 1 | 0.99 | 11 | 1 | 0.99 | 20 | 3 | 0.98 |
| 3 | 1 | 0.98 | 12 | 1 | 0.99 | 21 | 5 | 0.98 |
| 4 | 1 | 0.98 | 13 | 1 | 0.98 | 22 | 4 | 0.97 |
| 5 | 1 | 0.97 | 14 | 2 | 0.98 | 23 | 4 | 0.97 |
| 6 | 1 | 0.97 | 15 | 3 | 0.97 | 24 | 3 | 0.96 |
| 7 | 1 | 0.96 | 16 | 1 | 0.98 | 25 | 3 | 0.96 |
| 8 | 2 | 0.96 | 17 | 2 | 0.98 | 26 | 5 | 0.96 |
| 9 | 3 | 0.96 | 18 | 2 | 0.98 | | | |

| Conductor Cost = $3325.6 | Loss Cost = $ 5961.21 |
|---|---|
| $U_{ind}$ = 0.4939 | $\sum_{t=1}^{T} Ploss_t$ = 1752.6 kW |
| Total_Cost = $9286.86 | |

These results of this scenario, named *case 1*, show the best possible arrangement of conductors for minimizing the system's costs. In Table IV and all the subsequent tables, *Total_Cost* is the same as *OBJ* defined in (1). As seen, by approaching the end of the feeders, smaller conductor sizes have been selected. This is because in this system, the current flows decrease by approaching the end of the feeders.

In order to analyze the effect of each component of *OBJ*, i.e. *Cond_Cost* and *Loss_Cost* on the obtained results, another objective function as (28), in which the cost components are assigned weights, is defined to study the effect of each cost component in the conductor arrangement.

$$OBJ^* = \omega \times \sum_{k=1}^{N_k} Cond\_Cost_k + (1-\omega) \times \sum_{t=1}^{T} Loss\_Cost_t \quad (28)$$

By changing the weighting factor, $\omega$, in small steps from 0 to 1, the planning problem is solved several more times. The results are shown in Fig. 2 and Fig. 3.

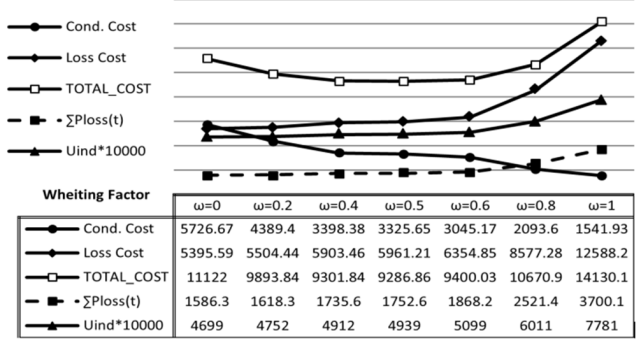

| Wheiting Factor | $\omega=0$ | $\omega=0.2$ | $\omega=0.4$ | $\omega=0.5$ | $\omega=0.6$ | $\omega=0.8$ | $\omega=1$ |
|---|---|---|---|---|---|---|---|
| Cond. Cost | 5726.67 | 4389.4 | 3398.38 | 3325.65 | 3045.17 | 2093.6 | 1541.93 |
| Loss Cost | 5395.59 | 5504.44 | 5903.46 | 5961.21 | 6354.85 | 8577.28 | 12588.2 |
| TOTAL_COST | 11122 | 9893.84 | 9301.84 | 9286.86 | 9400.03 | 10670.9 | 14130.1 |
| $\sum Ploss(t)$ | 1586.3 | 1618.3 | 1735.6 | 1752.6 | 1868.2 | 2521.4 | 3700.1 |
| $U_{ind}*10000$ | 4699 | 4752 | 4912 | 4939 | 5099 | 6011 | 7781 |

Figure 2. The Results of *OBJ\** for $\omega$ changing from 0 to 1

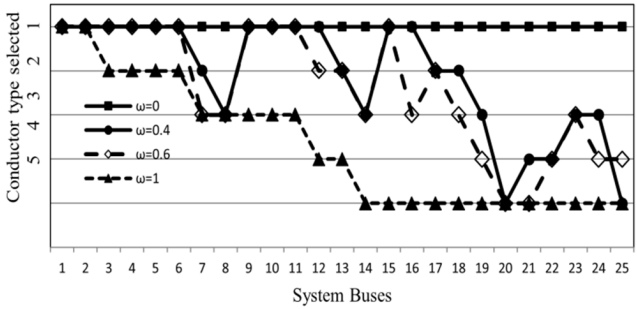

Figure 3. Conductor profile for different values of $\omega$ in *OBJ\**

The case with $\omega=0$ corresponds to minimizing the system's loss cost only. Thus, as seen from the conductor profile of Fig. 3, the PSO has chosen the conductors of the least resistance, i.e. *type 1* for all sections which leads to the highest conductor cost, as shown in Fig. 2. Figure 2 shows that the lost power, loss cost, and $U_{ind}$ are at their minimum values for this case. With the values of ω being gradually increased, the weight of the conductor cost rises. Thus, the optimization problem select smaller size conductors compared to the case with ω =0. Therefore, the values of the lost power, loss cost, and $U_{ind}$ are gradually increased and the conductor cost decreases. Finally in the case with ω=1, corresponding to the conductor cost minimization only, the least expensive conductors are chosen for the system, as seen in the conductor profile of Fig. 3, considering allowable voltage and current limits. This leads to the highest loss cost and voltage drop throughout the system as shown in Fig. 2. However, the optimal composition of the conductors is achieved when the system's total cost is at its minimum value which occurs at ω=0.5, i.e. the same as *case1*. This corresponds to the equal consideration of the two cost components.

***Scenario B.*** *Conductor sizing along with Capacitor and DG placement:* It is assumed that a total budget of $5000 and $10000 are allocated for capacitors and DGs placement as (16) and (17), respectively. The optimal planning of all the problem components, simultaneously, is investigated in a new case named *case 2*. Table V presents the results of this case. The thorough change in the conductors' arrangement and considerable decrease in *Conductor Cost* and *Total_Cost* compared to those of *case 1* is observed. Also, the value of $U_{ind}$ and total power loss are highly improved compared to those of *case 1*. As expected, in presence of DGs and capacitors, due to the current flow decline through the system branches, some of the conductors are replaced with weaker ones. For example, it can be inferred that because of the DG selected to be installed at bus 25, both active and reactive current flow and subsequently power loss has decreased in section 25. Hence, the proposed PSO algorithm has chosen a weaker conductor for this section compared to the *case 1* to achieve a more economic result. This observation can also be made for other buses with installed capacitors and DGs. Besides, it is seen that an amount of $4980 out of the total $5000, which is the maximum possible value with the given capacitors' data, is spent for the three chosen capacitors. Figure 4 shows the final resulted system arrangement of *case 2*, schematically.

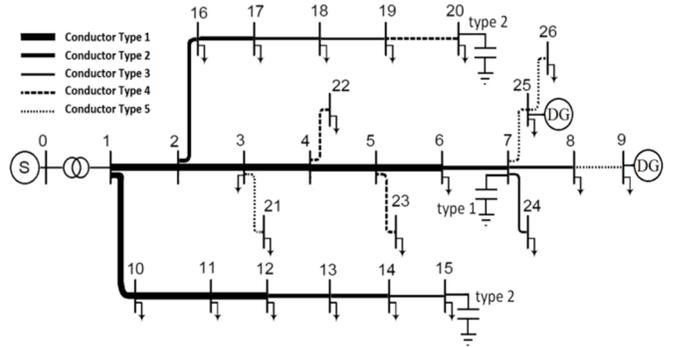

Figure 4. Final components' arrangement for the test system (*case 2*)

## V. CONCLUSION

Distribution system planning, including conductor sizing along with capacitor and DG sizing and placement, is investigated in this paper. The aim of minimizing the summation of loss cost and conductor cost is pursued. The results prove that the capacitors' and DGs' places change the conductors' optimal disposition dramatically. Besides, by assigning different weights to the objective function's components, the conductor sizing sub-problem is studied and the results are analyzed and compared. An innovative method to give PSO the ability of selection from standard values is proposed in this paper which is shown to be capable of handling both binary and selective variables, effectively.

TABLE V. THE SIMULTANEOUS CONDUCTOR, CAPACITOR, AND DG ARRANGEMENT RESULTS OF CASE2

| \multicolumn{8}{c}{Conductors Arrangement} | \multicolumn{2}{c}{Capacitors Arrangement} | \multicolumn{2}{c}{DG Arrangement} |
|---|---|---|---|---|---|---|---|---|---|---|---|
| Sec. | Cond. | Sec. | Cond. | Sec. | Cond. | Sec. | Cond. | Bus | Cap. | Bus | DG |
| 1 | - | 8 | 3 | 15 | 3 | 22 | 4 | 7 | 1 (1200 kVAr) | 9 | 500kW- 300kVAr |
| 2 | 1 | 9 | 5 | 16 | 2 | 23 | 4 | 15 | 2 (600 kVAr) | 25 | 500kW- 300kVAr |
| 3 | 1 | 10 | 1 | 17 | 2 | 24 | 3 | 20 | 2 (600 kVAr) | | |
| 4 | 1 | 11 | 1 | 18 | 3 | 25 | 5 | | | | |
| 5 | 1 | 12 | 1 | 19 | 3 | 26 | 5 | Capacitor Cost = $4980 | | DG Cost = $8000 | |
| 6 | 1 | 13 | 2 | 20 | 4 | | | Conductor Cost = $2706.03 | | Loss Cost = $4010.56 | |
| 7 | 2 | 14 | 2 | 21 | 5 | | | | | | |

$\sum_{t=1}^{T} Ploss_t = 1174.9 \text{ kW}$, $\quad Uind = 0.3713$, $\quad Total\_Cost = \$6716.59$